# Modeling electrostatic and quantum detection of molecules

S. Vasudevan[1], K. Walczak[1], N. Kapur[2], M. Neurock[2], A.W. Ghosh[1]

[1] Department of Electrical and Computer Engineering, University of Virginia
Charlottesville, Virginia 22904, USA
[2] Department of Chemical Engineering, University of Virginia
Charlottesville, Virginia 22904, USA

We describe two different modes for electronically detecting an adsorbed molecule using a nanoscale transistor. The attachment of an ionic molecular target shifts the threshold voltage through modulation of the depletion layer electrostatics. A stronger bonding between the molecule and the channel, involving actual overlap of their quantum mechanical wavefunctions, leads to scattering by the molecular traps that creates characteristic fingerprints when scanned with a backgate. We describe a theoretical approach to model these transport characteristics.

## I. INTRODUCTION

The development of electronic 'touch' sensors has far-reaching implications for the real-time, reagentless detection of single-molecules using a scalable, solid state platform. The prototype electronic sensor device is a ChemFET (Chemically modulated Field Effect Transistor), which works on purely electrostatic principles. In such a sensor, an ionic target modulates the accumulation or depletion layer charges and thereby the channel resistance of a receptor functionalized, back-gated silicon-on-insulator (SOI) or a nanowire transistor [1]. The attachment of a target molecule transfers charge to the receptor and the resulting electronic repulsion shifts the corresponding transistor threshold voltage (Fig.1a). Although its operational principle is quite simple, the device has some limitations – it relies on the detection of a single quantity, the attached charge, making it difficult to detect an unknown molecular species. The long-range nature of the electrostatic signals makes the device prone to false positives created by stray charges in its vicinity. Parallel detection of many species of molecules will require creating a parallel array of variously functionalized wires with accompanying challenges related to alignment and cross-talk. Finally, the threshold shift needs to be completely immune to the dynamics of traps at the silicon-oxide interface.

    A radically different principle of operation involves chemical bonding with a molecular species, transferring charge and spectral weight between the molecule and the silicon surface either through direct attachment, through an intervening oxide tunnel barrier or through a receptor that provides a convenient superexchange pathway. The overlap of molecular and silicon wavefunctions serves to passivate existing surface states as well as to create new localized molecular trap levels inside the silicon band-gap. At resonance driven by a gate, the traps are stochastically filled and emptied by the channel electrons [2], blocking and unblocking the channel through long-ranged electrostatic repulsion (Coulomb scattering) as well as short-ranged quantum interference (Fano scattering) [3]. The resulting two-state random telegraph signal (RTS) can be used to locate the trap position both spectrally as well as spatially [4,5]. Such principles of trap detection are short ranged and have the potential of being truly selective, with a large signal-to-noise ratio [6]. The effect can be enhanced in modern nanodevices as they can be fabricated practically defect free with near ballistic levels

of operation. In contrast with ChemFETs, where one detects a single threshold shift for a specific molecule, here we encounter an entire spectral nano-'barcode' that can be compared against a compiled table of theoretical responses to characterize and sense a molecular species. Since these devices operate by modulating surface properties of transistors, we call them 'SurfFETs'. The significant advantage of such SurfFETs is their exclusive detection of only molecules that overlap significantly with the channel to cause a transfer of states. Remote, physisorbed molecules in the vicinity do not create traps for RTS detection. This paper develops the basic model behind charge detection in ChemFETs and SurfFETs (Fig.1).

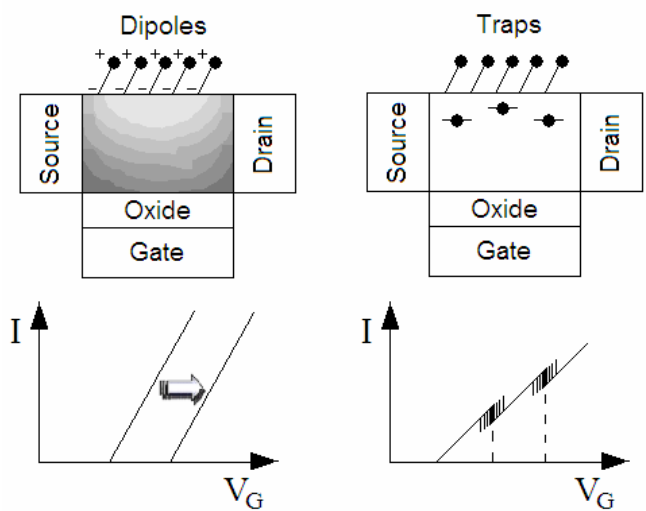

*Figure 1*: Schematic figure of the sensor geometry and dipole-induced threshold voltage shift in a ChemFET (left), and trap-induced random telegraph signals in a SurfFET (right).

There has been little modeling effort in quantitative understanding of electronic sensing, principally due to the complex geometry and chemistry at the molecule-receptor and the receptor-transistor interfaces, as well as the interfacial electronic structure that combines both short-ranged chemical and long-range band correlations. For a ChemFET, the principal evolution equations are simple, and the difficulty is with the practical implementation that will need to combine atomistic charge-transfer processes with macroscale electrostatics. For FETs long enough to have slowly varying channel potentials, the macroscopic band-bending and the microscopic charge transfer processes can be treated independently. We can thus use Density functional theory (DFT) to compute the reconstructed silicon and molecular attachment geometries, the interfacial dipole and band-alignment, and incorporate these results into a separate Poisson solver to compute the band-bending along the transistor depth direction. Dipole-induced threshold voltage shifts calculated using this approach are in good agreement with recent experimental data [7,8]. Modeling quantum detection in SurfFETs is more complicated and needs attention to many-body effects in the molecules that act as quantum dots with strong confinement. Capturing these many-body excitations quantitatively will require solving a multi-electron master equation for Coulomb Blockade and multiphonon processes in the dot [9-11]. Since the underlying transistor channel electrons are weakly interacting, we can then extract a time-dependent 'mean-field' or one-electron response function that can be used to simulate Coulomb or quantum scattering by the dot. The temporal response is then incorporated through a potential or a self-energy function into a time-dependent non-equilibrium Green's function (TDNEGF) formalism [3,12-14] to calculate the voltage sweep-dependent channel current.

## II. ELECTROSTATIC DETECTION OF MOLECULES (ChemFETs)

Let us first discuss the surface modulation of transport characteristics in a ChemFET, where dipolar molecules grafted on the conduction channel affect the threshold voltage of the conventional Field-Effect Transistor (FET) (Fig.1a). In the diffusive transport regime for the FET, the transistor current-voltage (I-V) characteristics follow standard square law theory [15]:

$$I = \frac{eC_{ox}\mu W}{2L}\left[V_G - V_T - m\frac{V}{2}\right]V, \quad (1)$$

where $W$ and $L$ are the channel width and length, $C_{ox}$ is the oxide capacitance, $\mu$ is the channel mobility, $V_G$, $V_T$ and $V$ are the gate voltage, threshold voltage and source-drain bias respectively, and $m = 1 + (t_{ox}/t_{ch})(\varepsilon_{ch}/\varepsilon_{ox})$ is the body correction factor, $\varepsilon$ denoting the dielectric constant and $t$ denoting the thickness of the oxide ($ox$) and the channel ($ch$). As schematically illustrated in Fig.2, we use density functional theory (DFT) within the local density approximation with gradient corrections (LDA-GGA) in the Vienna Ab-Initio Simulation Package (VASP) [16,17], to obtain the reconstructed geometry at the silicon-molecule interface. The method also gives us the shift in workfunction and the interfacial dipole, which are then included into a Poisson solver in ADEPT [18] to calculate the band-bending and shift in threshold voltage, as shown schematically in Fig.2.

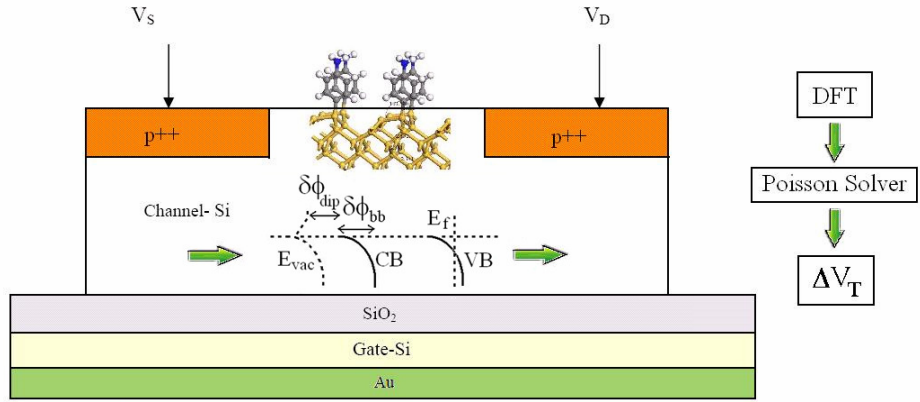

*Figure 2*: *Atomistic simulation of molecular attachment and interfacial dipole, computed with DFT, is used to compute the Poisson potential and band-bending of the ChemFET and hence its threshold voltage shift.*

There are two contributions to the threshold voltage shift in a ChemFET. The purely dipolar components give a shift in potential given by

$$\delta\phi_{dip} = N_{mol}\left(\frac{\mu_{dip}}{\varepsilon_m}\right)\cos(\theta)e^{-t_{ch}/L_D}, \quad (2)$$

where $N_{mol}$ gives the areal surface density of the attached molecules, $\mu_{dip}$ is the static dipole moment of their headgroups oriented at an average angle $\theta$ relative to the normal, $\varepsilon_m$ is the molecular dielectric constant, and $t_{ch}$ is the channel thickness to its backgate. Electrostatic screening is characterized by the Debye length $L_D = \sqrt{\varepsilon_r \varepsilon_0 k_B T / ne^2}$, where $\varepsilon_r$ is the dielectric constant of the channel, $\varepsilon_0$ is the permittivity of free space, $T$ is the temperature and $n$ is the channel electron density. As expected, the contribution is opposite for opposite dipolar signs as the charges effectively gate the channel and lead to further accumulation or depletion of charge.

The computed dipolar shifts for two distinct molecular species (nitrobenzene and aniline) [8] agree quantitatively with experimental values, obtained by combining the data from UPS/IPES and XPS measurements [19]. In addition, the experiment shows a larger contribution to the threshold shift which maybe due to band-alignment with the uncharged molecular backbone that leads to transfer of charge to the transistor top surface. The shift is given by the difference between the channel Fermi energy $E_F$ (which is often pinned by surface states) and the charge neutrality potential $E_{CNL}$ of the molecular backbone (the energy to which electrons need to fill the molecule to keep it electrically neutral). The work-function shift of the molecule is then given by

$$\delta\phi_{bb}(mol) = \frac{E_F - E_{CNL}}{1 + 1/U_0 D_0}, \tag{3}$$

where $U_0$ is the single-electron charging energy of the molecular level and $D_0$ is its density of states near the Fermi energy. Assuming the molecular states are well tied with the silicon surface, this will also shift the surface potential of the channel by the same amount, which gets transferred to the bottom of the channel, through the same exponential Debye screening factor, so that $\delta\phi_{bb} = \delta\phi_{bb}(mol)e^{-t_{ch}/L_D}$. The final shift in threshold voltage is then given by

$$\Delta V_T = \delta\phi_{dip} + \delta\phi_{bb}. \tag{4}$$

The computed shifts due to the dipole and the charge transfer for the different systems have been tabulated in Table 1 and agree quantitatively with experiments [7].

**Table 1:** *Data of computed dipoles of the molecule-semiconductor system, shifts due to dipole ($\delta\phi_{dip}$) and charge transfer ($\delta\phi_{bb}$), computed and experimental changes [7] in the threshold voltage of pseudo-MOSFETs. The control consists of an H-passivated Si(100) surface.*

| Systems | Dipole Moment (dB) | $\delta\phi_{dip}$ (V) | $\delta\phi_{bb}$ (V) | $\Delta V_T(V)$ Theory | $\Delta V_T(V)$ Expt. |
|---|---|---|---|---|---|
| Control | 0.03 | 0.0058 | 0 | 0.0058 | 0 |
| Nitrobenzene-Silicon | 2.17 | 0.22 | -0.725 | -0.525 | -0.55 |
| Aniline-Silicon | -1.44 | -0.14 | -0.96 | -1.1 | -1.25 |

The experiments were performed on dipolar molecules with a low-doped pseudoMOSFET, providing a very long Debye length (~ 1 µm for a doping level of $10^{13}$ cm$^{-3}$) that allowed the threshold voltage of the backgated pseudoMOSFET to be sensitive to dipolar fields from their surface. In these devices, the current spreads beyond the bottom layer as the FET is operating in the accumulation rather than the inversion mode. Our calculations in Table I were calibrated to these specific experimental results. For reliable device-level integration, however, the FETs need to be doped strongly to make them robust with respect to variation of device parameters. Since this would shorten the Debye length substantially, this would require thinning down the channel considerably, such as by employing silicon nanowire FET sensors, for example [20]. The approach outlined in this paper is quite general and carries over to such FETs as well, although the details of the molecular attachment on the nanowires (necessitating a DFT-based computation) as well as the screening of the dipolar electrostatics by the lower-dimensional wire will be different. In addition, for nanowire FETs, the square-law theory needs to be modified [21]. The underlying quasi-ballistic equations can be obtained by replacing the oxide capacitance with the semiconductor quantum capacitance, and the mobility limited electron velocity by a harmonic mean of the the contact injection velocity and the band velocity of the channel material.

### III. QUANTUM DETECTION OF MOLECULES (SurfFETs)

The operating principle for a SurfFET is the creation of molecule-specific traps, whose stochastic filling/emptying near resonance alternately blocks and unblocks the underlying transistor channel and creates a corresponding flicker in its output current. The ensemble of gate voltage windows over which such random telegraph signals (RTS) manifest themselves map out the molecular 'bar-code', thus functioning as a powerful characterization tool. RTS has already been successfully employed to detect a single $P_b$ center dangling bond in a commercial transistor [22], and also to detect discrete defects in carbon nanotubes [6]. The process can be combined with other probe techniques (temperature, magnetic field, etc) to further characterize these systems more precisely.

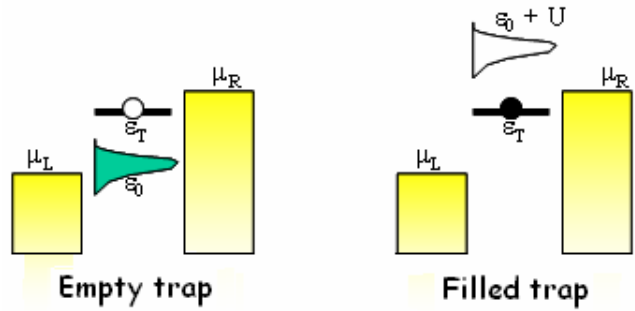

*Figure 3: A gate voltage can be used to move the trap level $\varepsilon_T$ into resonance with the contact Fermi energies, alternately filling and emptying the trap. Filling an initially empty trap pushes the channel level from $\varepsilon_0$ to $\varepsilon_0 + U$ by Coulomb repulsion, expelling it from the conducting window and turning off the current. The shaded regions show the contact states filled up to their respective electrochemical potentials $\mu_{L,R}$.*

The simulation of RTS dynamics would, in general, require coupling the response of a strongly correlated quantum dot, solved using a multielectron master equation [9-11], with a time-dependent non-equilibrium Green's function (TDNEGF) formulation [3,12-14] for electron transport in the weakly interacting transistor channel. For the sake of simplicity, we will adopt a simpler model where all transistor modes act as independently conducting channels. We will also ignore 'memory' effects by invoking an adiabatic approximation, simply making the input parameters of the steady-state NEGF transport equations [23] time-dependent. This can be justified since the electron capture and emission rates in the channel must be much slower than the voltage sweep rate in order for their detection. Under these circumstances, the steady-state current at the left contact is written using the Landauer formula [24]:

$$I_L = \frac{2e}{h}\int_{-\infty}^{+\infty} d\varepsilon T(\varepsilon)[f_L(\varepsilon) - f_R(\varepsilon)], \qquad (5)$$

where $f_{L,R}(\varepsilon)$ are the bias-separated contact Fermi functions $f_\alpha(\varepsilon) = (\varepsilon - \mu_\alpha)$ for the $\alpha = L, R$ electrode, with electrochemical potentials $\mu_L = E_F$, $\mu_R = E_F + eV$ related to Fermi energy $E_F$. The transmission function $T(\varepsilon)$ is determined by the channel couplings (assumed equal, $\gamma_L = \gamma_R = \gamma$) with the source and drain contacts, and the channel Green's function $G_{ch}(\varepsilon)$ whose imaginary part determines its density of states $D_{ch}(\varepsilon)$:

$$T(\varepsilon) = \gamma^2 |G_{ch}(\varepsilon)|^2. \qquad (6)$$

The stochastic dynamics of the dot influences the channel Green's function through Coulomb repulsion, making it time-dependent. For an initially empty dot with a single energy level $\varepsilon_0$:

$$G_{ch}(\varepsilon,t) \cong \frac{1 - n_{dot}(t)}{\varepsilon - \varepsilon_0 + i\gamma} + \frac{n_{dot}(t)}{\varepsilon - \varepsilon_0 - U + i\gamma}. \qquad (7)$$

The dot occupancy $n_{dot}(t)$ varies stochastically between 1 and 0 near resonance (single-electron limit, assuming a weak dot-channel coupling). Since the trap itself is non-conducting, filling the trap kicks the formerly conducting channel level $\varepsilon_0$ out of the source-drain bias window through Coulomb repulsion $U$, thus blocking the channel and reducing its current. Subsequently emptying the trap unblocks it. We can easily generalize this approach to other channel densities of states. For instance, the Green's function for the inversion channel of a 2D MOSFET is given by:

$$G_{ch}(\varepsilon,t) \cong \frac{m^*S}{2\pi\hbar^2}[[1 - n_d(t)]\Theta(\varepsilon - E_C) + n_d(t)\Theta(\varepsilon - E_C - U)], \qquad (8)$$

where $m^*$ is the channel effective mass, $S$ is the injection area of the charge carriers, and $E_C$ denotes bottom of the conduction band, and $\Theta$ is the Heaviside step function.

The time-dependence of the dot occupancy over a time-step $dt$ can be extracted from a Monte Carlo simulation with a capture probability $dt/\tau_c$ and an emission probability $dt/\tau_e$, the capture and emission times given by:

$$\hbar/\tau_c = \gamma_s F(\varepsilon_T - \alpha_G V_G - \alpha_D V_D), \qquad (9a)$$

$$\hbar/\tau_e = \gamma_s [1 - F(\varepsilon_T - \alpha_G V_G - \alpha_D V_D)], \qquad (9b)$$

$$\gamma_s = 2\pi |\tau|^2 D_{ch}(\varepsilon), \qquad (9c)$$

where $D_{ch}(\varepsilon)$ is the local density of states of the channel at the site of the trap, $\tau$ is the quantum coupling or bond-strength between trap and channel, the dimensionless factors $\alpha_{G,D}$ denote the capacitive transfer factors between applied and local potentials at the gate and drain respectively, and $F$ is the occupancy function for the trap, assumed to be in equilibrium with the channel

$$F(\varepsilon) = \left[1 + \frac{1}{2}\exp\left[\frac{\varepsilon - E_F}{k_B T}\right]\right]^{-1}. \tag{10}$$

The factor $1/2$ represents the double-degeneracy of the paramagnetic dot state. The singlet state of the initially empty trap is assumed inaccessible due to its large on-site Coulomb cost.

The trap-channel interaction process is summarized pictorially in Fig.3, while the mathematical ingredients of our model are summarized in Fig.4. The method is readily generalizable to any channel density of states $D_{ch}(\varepsilon)$ beyond a single-level Lorenzian and 2D MOSFET assumed here. In the matrix TDNEGF form, it can include additional effects due to quantum interference and memory [3]. Through many-body rate equations, it can also include strong correlation effects on the dot. Finally, through additional self-energy matrices it can include dephasing events due to coupling with the environment.

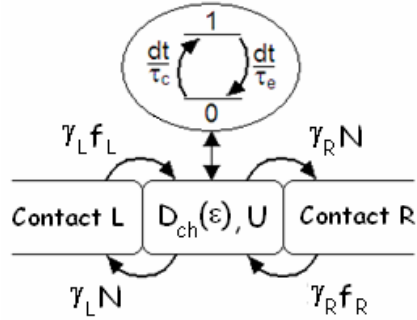

*Figure 4*: *RTS simulation approach consists of solving for the stochastic dot dynamics using Monte Carlo (Eqs.9,10) and using the output potential to shift the channel density of states (Eqs.7,8), which in turn is populated by the contacts at rates $\gamma_{L,R}$ to drive the current (Eqs.5,6).*

The trap-driven channel equations (5)-(10) explain the origin of RTS in nanoscale FETs. For a given gate voltage sweeping rate, the trap level $\varepsilon_T$ is shifted by the Laplace potential Eq.9, altering the corresponding capture and emission probabilities. Near resonance, the probabilities are almost equal, creating fast transitions between the two-states (0 and 1 in Fig.4), provided the voltage sweep rate is faster than the trap lifetime [25]. From the time-dependent current and a given voltage sweep rate, we can obtain its $I - V_g$, as shown in Fig.5. For suitably chosen parameters, our model can reproduce experimental data semi-quantitatively. Furthermore, the transitions among the various trap excitations are expected to be molecule-specific, leading to a corresponding uniqueness of the voltage windows over which the RTS signals are observed. By combining a trace of these windows, and their behavior under additional probes such as drain voltage or magnetic fields, we can characterize the sequence of excitations as well as the physical location of the trap along the channel.

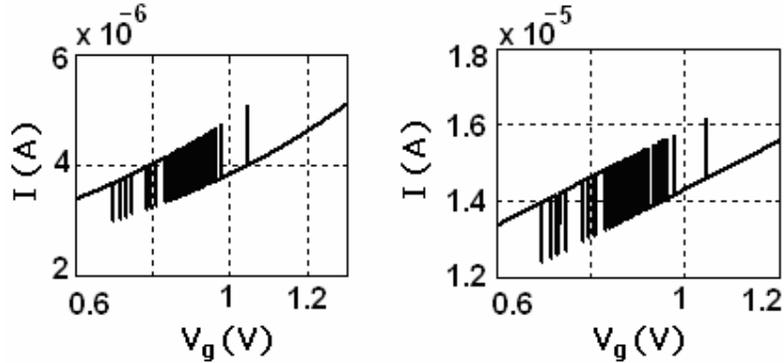

*Figure 5*: *Theoretical simulations for a channel with a single energy level using Eq.7 (left), and a two-dimensional density of states as in the inversion layer of an FET using Eq.8 (right). RTS persists over a window dictated by the applied drain bias. The switching between high and low current states starts with lower trap capture rates (predominantly high current states) at the start of the window and ends with higher capture rates (lower current states) once the trap nears full occupation*

## IV. CONCLUDING REMARKS

In contrast to traditional optical methods of molecular detection based on fluorescence assays, ChemFET or SurfFET based electronic detection can be integrated onto a scalable chip platform [26]. ChemFETs offer distinct advantages over the conventional ion-selective electrode [27], because of compactness, low-output impedance, fast response, and the ability of mass-fabrication. ChemFETs based on silicon nanowires, carbon nanotubes, and other nanomaterials have already proved to be very useful in detection of macromolecular complexes such as proteins, biomolecules and toxic gases [20,28-33]. The use of surface receptors has limitations, including the need for additional steps in an assay, the difficulty in detecting certain biochemical activities, and the inability to identify unanticipated molecules. The compatibility of the nanosensor fabrication technique with currently available complementary metal oxide semiconductor (CMOS) technology would allow the manufacturing of sensor arrays, complementary error detection, and integrated signal processing electronics [34,35]. Thus, ChemFETs enable direct, label-free, real-time, and continuous sensing of ionic or dipolar molecules, and are suitable for big molecular complexes. This nanosensor can also be useful in nanomedicine, which refers to highly specific medical inventions at the molecular scale, enabling detection of proteins and other biomolecules released by cancer cells. This is crucial from therapeutic point of view to diagnose cancer at its early-stage of development [36,37].

The fabrication of an RTS-based nanosensor is similar to a standard ChemFET from a lithographic point of view, although it needs a solid-state environment for Coulomb Blockade and specialized surface functionalization schemes to maximize wavefunction overlap with the channel. It should also provide adequate resolution of the scanned trap levels within the channel band-gap to allow selectivity of small chemical molecules with well-defined electronic structures. Detailed analysis of noise-like random telegraph signals, originating from only one or few impurities within the conduction channel of the transistor, will enable us to localize and characterize molecular impurities in the system. While RTS has been used extensively in the early days of transistor technology to weed out impurities, the novelty of its present-day incarnation arises from the small size of the underlying transistor that makes its transport quasi-ballistic and considerably more sensitive to surface processes, while enabling the deliberate engineering of few (often one or none) adsorbates, or ordered arrays of molecular adsorbates through self-assembly. Beyond its impact on the nanoscale sensing paradigm, the detection and characterization of surface impurities will be crucial for the future of transistor technology, as the aggressive scaling of electronic devices makes them sensitive to overwhelming amounts of low-frequency ($1/f$) noise arising from a thermodynamic superposition of RTS signals from uncontrolled dopants, defects and dangling bonds.

## V. ACKNOWLEDGEMENTS

We thank Keith Williams, Tsz Wah Chan, Brian Burke, Supriyo Datta, Ashraf Alam, John Bean, Lloyd Harriott and Lin Pu for useful discussions. We specially thank Keith Williams for contributing to the name 'SurfFETs'. This work was supported by DARPA-AFOSR, NSF-NIRT and the NSF Career Award.

## REFERENCES

[1] F. Patolsky, G. Zheng, and C.M. Lieber, "Nanowire-based Biosensors", Analyt. Chem., vol. 78, pp. 4260-4269, 2006.
[2] M. Xiao, I. Martin, and H.W. Jiang, "Probing the Spin State of a Single Electron Trap by Random Telegraph Signal", Phys. Rev. Lett., vol. 91, art. 078301, 2003.
[3] K. Walczak, S. Vasudevan, and A.W. Ghosh, "Theory of electronically addressed Rabi oscillations", unpublished, 2007.
[4] K.S. Ralls, W.J. Skocpol, L.D. Jackel, R.E. Howard, L.A. Fetter, R.W. Epworth, and D.M. Tennant, "Discrete Resistance Switching in Submicrometer Silicon Inversion Layers: Individual Interface Traps and Low Frequency (1/f?) Noise", Phys. Rev. Lett., vol. 52, pp. 228-231, 1984.
[5] D.H. Cobden, and M.J. Uren, "Random telegraph signals from liquid helium to room temperature", Microel. Eng., vol. 22, pp. 163-170, 1993.
[6] F. Liu, M. Bao, H. Kim, K.L. Wang, C. Li, X. Liu, and C. Zhou, "Giant random telegraph signals in the carbon nanotubes as a single defect probe", Appl. Phys. Lett., vol. 86, art. 163102, 2005.
[7] T. He, J. He, M. Lu, B. Chen, H. Pang, W.F. Reus, W.M. Nolte, D.P. Nackashi, P.D. Franzon, and J.M. Tour, "Controlled Modulation of Conductance in Silicon Devices by Molecular Monolayers", J. Am. Chem. Soc., vol. 128, pp. 14537-14541, 2006.
[8] S. Vasudevan, N. Kapur, and A.W. Ghosh, "Threshold voltage control in silicon transistors using molecular overlayers", unpublished, 2007.
[9] B. Muralidharan, A.W. Ghosh, and S. Datta, "Probing electronic excitations in molecular conduction", Phys. Rev. B, vol.73, art. 155410, 2006.
[10] B. Muralidharan, A.W. Ghosh, and S. Datta, "Conductance in Coulomb blockaded molecules: fingerprints of wave-particle duality?", Mol. Simul., vol. 32, pp. 751-758, 2006.
[11] B. Muralidharan, and S. Datta, "Generic model for current collapse in spin-blockaded transport", Phys. Rev. B, vol. 76, art. 035432, 2007.
[12] A.-P. Jauho, N.S. Wingreen, and Y. Meir, "Time-dependent transport in interacting and non-interacting resonant-tunneling systems", Phys. Rev. B, vol. 50, pp. 5528-5544, 1994.
[13] Y. Zhu, J. Maciejko, T. Ji, H. Guo, and J. Wang, "Time-dependent quantum transport: Direct analysis in the time domain", Phys. Rev. B, vol. 71, art. 075317, 2005.

[14] D. Hou, Y. He, X. Liu, J. Kang, J. Chen, and R. Han, "Time-dependent transport: Time domain recursively solving NEGF technique", Physica E, vol. 31, pp. 191-195, 2006.
[15] M.S. Sze, *Physics of Semiconductor Devices*, Wiley, New York, 1981.
[16] G. Kresse, and J. Furthmuller, "Efficient iterative schemes for ab initio total energy calculations using a plane-wave basis set", Phys. Rev. B, vol. 54, pp. 11169-11186, 1996.
[17] G. Kresse, and J. Furthmuller, "Efficiency of ab initio total energy calculations for metals and semiconductors using a plane-wave basis set", Comput. Mater. Sci., vol. 6, pp. 15-50, 1996.
[18] J.L. Gray, M. McLennan (2007), "ADEPT", http://www.nanohub.org/tools/adept/
[19] T. He, H. Ding, N. Peor, M. Lu, D.A. Corley, B. Chen, Y. Ofir, Y. Gao, S. Yitzchaik, and J.M. Tour, „Silicon/Molecule Interfacial Electronic Modifications", J. Am. Chem. Soc., vol. 130, pp. 1699 - 1710, 2008.
[20] Y. Cui, Q. Wei, H. Park, and C.M. Lieber, "Nanowire Nanosensors for Highly Sensitive and Selective Detection of Biological and Chemical Species", Science, vol. 293, pp. 1289-1292, 2001.
[21] J. Guo, J. Wang, E. Polizzi, S. Datta, and M. Lundstrom, "Electrostatics of Nanowire Transistors", IEEE Trans. Nanotech., vol. 2, pp. 329-334, 2003.
[22] M. Xiao, I. Martin, E. Yablonovitch, and H.W. Jiang, "Electrical detection of the spin resonance of a single electron in a silicon field-effect transistor", Nature, vol. 430, pp. 435-439, 2004.
[23] S. Datta, *Quantum Transport: Atom to Transistor*, Cambridge University Press, 2005.
[24] Y. Meir, N.S. Wingreen, "Landauer Formula for the Current through an Interacting Electron Region", Phys. Rev. Lett., vol. 68, pp. 2512-2515, 1992.
[25] M.-H. Tsai, H. Muto, T.P. Ma, "Random telegraph signals arising from fast interface states in metal-SiO2-Si transistors", Appl. Phys. Lett., vol. 61, pp. 1691-1693, 1992.
[26] S.R. Nicewarner-Pena, F. Griffth-Freeman, B.D. Reiss, L. He, D.J. Pena, I.D. Walton, R. Cromer, C.D. Keating, and M.J. Natan, "Submicrometer Metallic Barcodes", Science, vol. 294, pp. 137-141, 2001.
[27] M.J. Madou, and R. Cubicciotti, "Scaling issues in chemical and biological sensors", Proc. IEEE, vol. 91, pp. 830-838, 2003.
[28] E. Souteyrand, J.P. Cloarec, J.R. Martin, C. Wilson, I. Lawrence, S. Mikkelsen, and M.F. Lawrence, "Direct Detection of the Hybridization of Synthetic Homo-Oligomer DNA Sequences by Field Effect", J. Phys. Chem. B, vol. 101, pp. 2980-2985, 1997.
[29] S. Park, T.A. Taton, C.A. Mirkin, "Array-Based Electrical Detection of DNA with Nanoparticle Probes", Science, vol. 295, pp. 1503-1506, 2002.
[30] R.J. Chen, S. Bangsaruntip, K.A. Drouvalakis, N.W.S. Kam, M. Shim, Y. Li, W. Kim, P.J. Utz, and H. Dai, "Noncovalent functionalization of carbon nanotubes for highly specific electronic biosensors", Proc. Natl. Acad. Sci. USA, vol. 100, pp. 4984-4989, 2003.
[31] A. Star, J.-C. Gabriel, K. Bradley, and G. Gruner, "Electronic Detection of Specific Protein Binding Using Nanotube FET Devices", Nano Lett., vol. 3, pp. 459-463, 2003.
[32] A.D. McFarland, and R.P. Van Duyne, "Single Silver Nanoparticles as Real-Time Optical Sensors with Zeptomole Sensitivity", Nano Lett., vol. 3, pp. 1057-1062, 2003.
[33] T. Tang, X. Liu, C. Li, B. Lei, D. Zhang, M. Rouhanizadeh, T. Hsiai, and C. Zhou, "Complementary response of In2O3 nanowires and carbon nanotubes to low-density lipoprotein chemical gating", Appl. Phys. Lett., vol. 86, art. 103903, 2005.
[34] J. Hahn, and C.M. Lieber, "Direct Ultrasensitive Electrical Detection of DNA and DNA Sequence Variations Using Nanowire Nanosensors", Nano Lett., vol. 4, pp. 51-54, 2004.
[35] E. Stern, J.F. Klemic, D.A. Routenberg, P.N. Wyrembak, D.B. Turner-Evans, A.D. Hamilton, D.A. LaVan, T.M. Fahmy, and M.A. Reed, "Label-free immunodetection with CMOS-compatible semiconducting nanowires", Nature, vol. 445, pp. 519-522, 2007.

[36] G. Zheng, F. Patolsky, Y. Cui, W.U. Wang, and C.M. Lieber, "Multiplexed electrical detection of cancer markers with nanowire sensor arrays", Nature Biotechnology, vol. 23, pp. 1294-1301, 2005.

[37] A.H. Ting, K.M. McGarvey, and S.B. Baylin, "The cancer epigenome-components and functional correlates", Genes Dev., vol. 20, pp. 3215-3231, 2006.